\documentstyle[fleqn,epsfig]{aipproc}

\begin{document}
\title{Physics Prospects with an \\ Intense Neutrino Experiment}

\author{Nickolas Solomey}
\address{Enrico Fermi Institute, The University of Chicago, Illinois 60637}

\maketitle

\begin{abstract}
With new forthcoming intense neutrino beams,
for the study of neutrino oscillations, it is possible to consider other
physics experiments that can be done with these extreme neutrino
fluxes available close to the source.
\end{abstract}

\section*{Introduction}
The objectives of my talk are to bring to the attention of both
the particle and nuclear physics communities the unique physics potential
that an intense neutrino beam experiment might be able to perform, other than 
that of neutrino oscillation which is covered elsewhere in these
proceedings \cite{james}. The list of non-oscillation
physics is large and will not be completely covered in detail in 
such a short writeup, but the hope is that it might interest physicists,
both theoretical and experimental, to contribute their ideas to
help formulate a new experiment that can take full advantage of the
intense neutrino beams planned in the NuMI project at Fermilab \cite{numi} and
eventually the muon collider prototype storage ring \cite{mucollider} which 
might be located at Fermilab along the NuMI beamline direction.

\section*{Physics Prospects}
Possible physics from an intense neutrino beam are:
\vspace{0.15cm}
\paragraph*{Scattering:}
The first such experiments were performed by
Rutherford with alpha scattering which showed the nuclei exists as a 
concentrated center of positive charge, while direct evidence of the quarks 
themselves came from experiments using the highest possible energy
electrons to probe the nucleon structure. Muon scattering experiments 
have also contributed similar results, as well as offered a more massive
probe. Furthermore, greatly improved statistics of electron
scattering reactions are now being done at CEBAF. The physics of these
experiments aim to measure structure functions with high precision. This is
done by measuring the electron scattering cross-section: 
\begin{equation}
\frac{d^2\sigma}{d\Omega dE'} = \frac{4\alpha^2(E')^2}{Q^4} 
cos^2(\theta/2) \times [\frac{F_2(x,Q^2)}{(E-E')} + 
\frac{2F_1(x,Q^2)}{M} tan^2(\theta/2)]
\end{equation}
Where $\alpha=1/137$, $M$ is the proton rest mass, $E$ is the electron 
energy before and $E'$ after the scattering, $\theta$ is the
electron scattering angle, and the Bjorken scaling variable is 
$x\equiv\frac{Q^2}{2M(E-E')}$. With neutrino scattering the 
expression changes
a little: $\frac{4\alpha^2(E')^2}{Q^4}$ becomes $\frac{G^2}{2\pi}$ where 
$G$ is the Fermi constant of weak coupling, and another structure 
function $F_3$ is needed because of parity violation. Neutrino 
scattering experiments are capable of measuring in detail the internal 
structure of the nucleon threw such experiments, with the advantage that 
the neutrino interactions are purely weak processes. They too can look 
towards further improvements with higher statistics. Eventually with a 
sufficiently large
data sample that can permit the data to be separated into $\nu$ and $\bar{\nu}$
this can be used to study the similarities between quarks and antiquarks.
\vspace{0.15cm}
\paragraph*{Strangeness Production:}
The most interesting new measurement would be 
strange particle production by neutrino interactions. This is suppressed
by a factor of $tan^2\theta_c$ (where $\theta_c$ is the Cabibbo mixing angle) 
and can only be done once an
intense neutrino source is available with an experiment capable of
identifying strange particles. With a small dedicated experiment of good
resolution, events can be seen of the charged current (CC)
and neutral current (NC) type:
\begin{eqnarray*}
\bar{\nu_l} + p^+ \rightarrow l^+ + \Lambda^0 \hspace{3cm}
& \bar{\nu} + p^+ \rightarrow \bar{\nu} + \Sigma^+ \hspace{1.5cm} SCNC \\ 
\rightarrow l^+ + \Sigma^0 \hspace{3cm} &  
\hspace{1.2cm} \rightarrow \bar{\nu} + \pi^0 + \Sigma^+ \hspace{0.6cm} SCNC \\ 
\rightarrow l^+ + \pi^0 + \Lambda^0 \hspace{2.1cm} &
\rightarrow \bar{\nu} + K^0 + \Sigma^+ \hspace{0.5cm} \\
\rightarrow l^+ + \pi^0 + \Sigma^0 \hspace{2.1cm} &
\hspace{1.3cm} \rightarrow \bar{\nu} + K^0 + p^+ \hspace{0.5cm} SCNC \\
\rightarrow l^+ + K^- + p^+ \hspace{1.9cm} & 
\hspace{1.3cm} \rightarrow \bar{\nu} + K^+ + n^0 \hspace{0.5cm} SCNC \\
\vdots \hspace{4.5cm} & \vdots \hspace{3cm} \\
\bar{\nu_l} + n^0 \rightarrow l^+ + \Sigma^- \hspace{2.8cm} &
\bar{\nu} + n^0 \rightarrow \bar{\nu} + \Lambda^0 \hspace{1.5cm} SCNC \\
\rightarrow l^+ + \pi^- + \Lambda^0 \hspace{1.9cm} &
\hspace{1.3cm} \rightarrow \bar{\nu} + \Sigma^0 \hspace{1.5cm} SCNC \\
\rightarrow l^+ + \pi^0 + \Sigma^- \hspace{1.9cm} & 
\hspace{0.05cm} \rightarrow \bar{\nu} + K^0 + \Lambda^0 \hspace{0.5cm} \\
\rightarrow l^+ + K^- + \Lambda^0 \hspace{1.75cm} &
\hspace{0.05cm} \rightarrow \bar{\nu} + K^0 + \Sigma^0 \hspace{0.5cm} \\
\vdots \hspace{4.5cm} & \vdots \hspace{2.75cm}
\end{eqnarray*} 
To date only a handful of experiments have measured a few of these reactions, 
all with bubble chambers where the particle interaction and secondary 
particles produced could be explicitly identified. The best results and only 
cross sections measured is from the CERN-PS Gargamelle bubble chamber
experiment \cite{gargamelle} with 15 events of $\Lambda^0$ CC 
production at 2$\times$10$^{-40}$~cm$^2$/nucleon; while 7 events exist
from the ZGS bubble chamber which includes one neutral current
strange particle production event \cite{zgs}. More recent
experiments of the 80s and 90s have not been
able to measure such interactions because they use large dense
detectors to increase their neutrino interaction rate. With the advent of
intense neutrino beams this can be over come by using a single thin 
target with a high precession experiment downstream to identify all of
the secondary particles. Not only would such an experiment be able to
measure strange particle production cross sections for charged and neutral
current, but the comparison of the two could give an independent measurement
of the weak mixing angle for a few channels, and is also possible to
improve the main interactions by explicitly removing these strange
particle production reactions from the sample. Table 1 lists the total 
neutrino and antineutrino interaction rate, as well as the expected rate of
$\Lambda^0$ production in the NuMI medium energy neutrino beam. Some of the NC
interactions are forbidden and if seen at a low level would
be an indication of strangeness changing neutral currents (SCNC) which
is a worthy physics objective. Also the production of hyperons can
permit a study of hyperon polarization when produced by neutrinos, the
polarization of hyperons in fixed target experiments by
a proton beam was originally a surprise and is still
theoretically unexplained \cite{pondrum}.
\begin{table}[h!]
\caption{Neutrino interactions expected per year in a solid target
2.5 cm thick and 1x1 m$^2$, or for liquid
target with a volume of 10 m$^3$. The Lambda yield is for the reaction
$\bar\nu_{\mu} + p \rightarrow \mu^+ + \Lambda^0$.}
\begin{tabular*}{4.35in}{llll}
target & $\nu$ interactions/year & $\bar\nu$ interactions/year &
            Lambda yield/year \\ \hline
Fe & 1.0 M & 0.25 M & 40 k \\
C  & 0.3 M & 75 k  & 12 k \\
W  & 2.5 M & 0.63 M & 110 k \\
H$_2$ & 2.2 M & 0.55 M & 90 k \\
D$_2$ & 5.0 M & 1.25 M & 250 k \\ \hline
\end{tabular*}
\label{rate}
\end{table}
 
Hyperon beta decays $A \rightarrow B \hspace{0.1cm} e^{-} \hspace{0.1cm}
\bar{\nu}_{e}  \hspace{0.1cm}$ are important to study for their
weak decay form-factors which give an understanding of their underlying
structure \cite{solomey}. 
In the V-A formulation the transition amplitude is:
\begin{equation}M = \frac{G}{\sqrt{2}} <B|J^{\lambda}|A>{\bar{u}_{e}} 
\gamma_{\lambda} (1 + \gamma_{5}) u_{\nu} \end{equation}
The V-A hadronic current can be written as:
\begin{eqnarray*} <B|J^{\lambda}|A> = {\cal C} \hspace{0.1cm} i \hspace{0.2cm}
\bar{u}(B) & [ & f_{1}(q^{2})\gamma^{\lambda}
+ f_{2}(q^{2}) \frac{\sigma^{\lambda \upsilon}\gamma_{\upsilon}}{M_{A}} +
f_{3}(q^{2}) \frac{q^{\lambda}}{M_{A}} + \\
& [ & g_{1}(q^{2}) \gamma^{\lambda} + g_{2}(q^{2}) 
\frac{\sigma^{\lambda \upsilon} \gamma_{\upsilon}}{M_{A}} + 
g_{3}(q^{2}) \frac{q^{\lambda}}{M_{A}} ]
\hspace{0.1cm} \gamma_{5} \hspace{0.25cm} ] u(A) 
\hspace{.1cm} (3) \end{eqnarray*}
where ${\cal C}$ is the CKM matrix element, and $q$ is the momentum transfer.
There are 3 vector form factors: $f_1$ (vector), $f_2$ (weak magnetism) and
$f_3$ (an induced scaler); plus 3 axial-vector form factors: $g_1$ (axial vector),
$g_2$ (weak electricity) and $g_3$ (an induced pseudo-scaler).
This is also possible to study with neutrino interactions that produce
hyperons, because the hyperon's decay itself has a self analyzing power
of the polarization. To obtain a large and clean data sample such measurements
can be unambiguous, avoiding the common problem of the missing neutrino 
which introduces multiple solutions and missing momentum.
\vspace{0.15cm}
\paragraph*{Comparison:}
The production of strangeness from nucleon
scattering using electrons or photons started in the 50s, but there is
yet no comprehensive theory \cite{david}. This data is modeled by many 
theorists, but the models only work for the
explicit interaction it was developed for. The final goal is to have
a description of all of the underlying reaction mechanisms
for strangeness production, and this is far from attained, especially since
the strangeness production by neutrino scattering are a grand total of only
26 events: 15 from one experiment,
7 from another and a few other experiments with one event each. However, 
an important message from the electron and photon 
experiments doing similar studies is the 
importance to investigate simultaneously all production channels \cite{saghai}:
\begin{eqnarray*}
\gamma + p^+ \rightarrow K^+ + \Lambda^0 \hspace{3cm} & 
e^- + p^+ \rightarrow e^- + K^+ + \Lambda^0 \\
\rightarrow K^+ + \Sigma^0 \hspace{3cm} &
\hspace{1.6cm} \rightarrow e^- + K^+ + \Sigma^0 \\
\rightarrow K^0 + \Sigma^+ \hspace{3cm} &
\hspace{1.6cm} \rightarrow e^- + K^0 + \Sigma^+ \\
\vdots \hspace{4.8cm} & \vdots \hspace{1.0cm}
\end{eqnarray*}
This is not a comprehensive list, other strange particle production
experiments have also been done with charged mesons: $\pi^{\pm}$ and 
$K^{\pm}$. Plus these reactions can be expanded to more complicated ones, 
but it is the simple processes that are most interesting. By extending these 
studies to include neutrino scattering experiments of strange particle 
production, it would give the ability to expand our understanding of the 
mechanism, aid in improving the underlying nucleon, hyperon and strange 
meson structure, and the best improvement will be the ability to compare 
what is found here with that of electromagnetic interactions. This is 
promising to expand our knowledge of the quark sea in the nucleon.
\vspace{0.15cm}
\paragraph*{Target Dependency:}
The most elementary particle structure information will come from
the simplest targets which is the lightest elements: hydrogen. While
neutrino scattering off of the atomic shell electron with high
statistics would be useful with both CC and NC interactions. However,
slightly more complicated reactions with
deuterium, tritium, $^3$He and $^4$He are also useful. For
example the deuterium data could be combined with the hydrogen
results to yield the neutron scattering information. While the
tritium and $^3$He data comparison provides information regarding
the rule of an extra proton or neutron in the nucleus of the simpler
nuclei. By using
heavier targets, which give higher rates, it is possible to study
the A dependency (number of nucleons, protons and neutrons, in the
target nuclei), $e^-$ and $\mu^{\pm}$ scattering experiments 
have found some intriguing results that show the quarks in heavy
nuclei are not simply confined to their protons or neutrons as some
models have suggested. Testing this with high statistics neutrino scattering 
experiments is important. This data could also be used to compare with nuclear
models that exist for N-N, $\pi$-N, e-N, $\gamma$-N and $\mu$-N
interactions; and eventually to formulate models of $\nu$-N and
$\bar{\nu}$-N interactions. With the ultimate goal of formulating
a comprehensive theory that works for all interactions simultaneously.
\vspace{0.15cm}
\paragraph*{$\tau$ Neutrino:}
With the high energy neutrino beam option of NuMI there is the possibility
to do $\nu_{\tau}$ detection from a point source which would further 
require a small emulsion active target, but due to particle
rates it would have to be changed often. The physics prospects are higher
statistics $\nu_{\tau}$ detection, and short baseline neutrino oscillation. 
In this case the detector would be used to trigger on those events to search
for in the emulsion. Due to the intense neutrino beams these emulsion
targets should be highly segmented and easily changed. The multiple drawer
system envisioned for the MINOS far detector is a possible choice
\cite{stanford}. If CP violation studies are ever to be done in the lepton 
sector, then the
oscillation rate of $\bar{\nu}_{\mu} \rightarrow \bar{\nu}_{\tau}$ from
disappearance will have to be compared to $\bar{\nu}_{\tau} \rightarrow
\bar{\nu}_{\mu}$ from appearance, but it will be necessary to have a small 
emulsion experiment at the near site to determine the $\bar{\nu}_{\tau}$ 
initial intensity.

\section*{Detector Requirements}
\begin{figure}[h]
\begin{center}
\mbox{
\epsfig{file=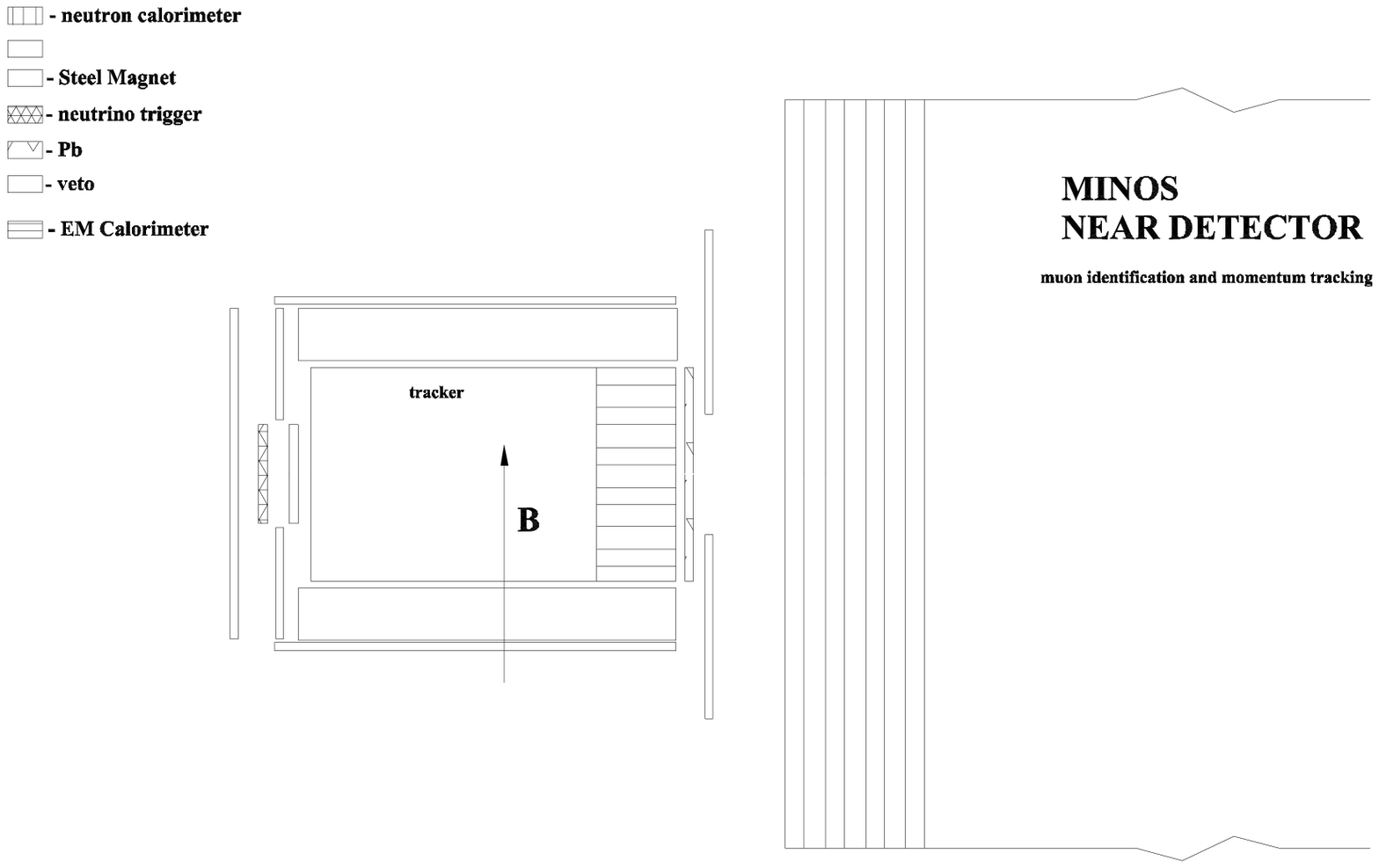,%
height=0.45\linewidth}
}
\end{center}
\caption{Conceptual design of a neutrino strange-particle 
production experiment.}
\label{scnc}
\end{figure}

An experiment to look for strange particle production by neutrino 
interaction does not have to be very large, but does have several
things it must perform well. A simple basic sketch of the 
conceptional detector design is shown in figure \ref{scnc}.
There is the need for a veto array in front of
the target and around the sides to catch any escaping particles,
a target which could possible be: liquid, different materials or maybe
even active. A tracking chamber in a magnetic field to be able to
track and measure the momentum of all the charged particles, with the
ability to reconstruct the decay vertex of the neutral strange 
particles ($\Lambda^0$,
$\Sigma^0$, K$^0$ ...). A good electromagnetic calorimeter, for electrons
and especially photon pairs to be able to see $\pi^0 \rightarrow \gamma 
\gamma$. There is the need for the ability to distinguish protons, $\pi^{\pm}$,
and K$^{\pm}$. Since the neutrino beam is not extremely high energy these
particles will be relatively low in momentum. Permitting their identification 
by dE/dx in the tracking chamber or TOF. 
By having the experiment sit in front of the MINOS near
detector, information from it can be used to identify muons and measure
their charge and momentum. Also, the first 3-5 interaction lengths in
the MINOS near detector can make a nice hadron calorimeter for identifying
neutrons.

There are many advantages for both experiments to be located together.
The cost effectiveness of producing the maximal physics from the 
NuMI neutrino beam is the most obvious. Each experiment can use information 
from the other to help calibrate. This experiment could measure more 
precisely the neutrino beam content and its radial distribution. 
Also protons and $\pi^-$ from $\Lambda^0$
decay can be identified and then the response of the MINOS near detector
studied so that hadron/muon identification in MINOS can be better understood
as a function of momentum.

\section*{Conclusions}
It is hoped that such an experiment can be organized in the next year
with detailed detector designs, physics goals achievable and a 
sufficiently funded collaboration formed. Such an experiment can be
of diverse interest to both the high energy and nuclear physics communities 
and represents an opportunity of cooperation. Furthermore, it will give
another reason for the NuMI project to operate intense neutrino beams,
which increase the physics yield from the neutrino flux that will be provided
for the other experiment. 

I would like to acknowledge discussion with Paul Schoessow, Amol Dighe,
Philippe Mine, Malcolm Derrick and Tom Fields.

\end{document}